\documentclass[final,1p,times,twocolumn]{elsarticle}

\usepackage{amssymb}
\usepackage{xcolor}
\usepackage{xurl}
\usepackage{float}

\usepackage{tabularx}
\usepackage{xcolor}
\usepackage{multirow}
\usepackage{csquotes}
\usepackage{longtable, booktabs}
\usepackage{siunitx}
\usepackage{array}
\usepackage{dcolumn}
\usepackage{pdfpages}
\usepackage{color, colortbl}
\usepackage[first=0,last=9]{lcg}

\definecolor{LightCyan}{rgb}{0.88,0.88,0.88}

\journal{Journal}

\begin{document}

\begin{frontmatter}



\title{Chemical strategies to mitigate electrostatic charging during coffee grinding}


\author[inst1]{Joshua M\'endez Harper}

\affiliation[inst1]{organization={Electrical and Computer Engineering, Portland State University},
            addressline={1900  SW 4th Avenue}, 
            city={Portland},
            postcode={97201}, 
            state={Oregon},
            country={US}; joshua.mendez@pdx.edu}

\author[inst2]{Christopher H. Hendon}
\affiliation[inst2]{organization={Department of Chemistry and Biochemistry, University of Oregon},
            addressline={1253 University of Oregon}, 
            city={Eugene},
            postcode={97403}, 
            state={Oregon},
            country={US}; chendon@uoregon.edu}

\begin{abstract}
The process of grinding coffee generates particles with high levels of electrostatic charge, causing a number of detrimental effects including clumping, particle dispersal, and spark discharge. At the brewing level, electrostatic aggregation between particles affects liquid-solid accessibility, leading to variable extraction quality. In this study, we quantify the effectiveness of four charge mitigation strategies. Our data suggests that adding small amounts of water to whole beans pre-grinding, or bombarding the grounds with ions produced from a high-voltage ionizer, are capable of de-electrifying the granular flows. While these techniques helped reduce visible mess, only the static reduction through water inclusion was found to impact the brewing parameters in espresso format coffee. There, wetting coffee with less than 0.05 mL / g resulted in a marked shift in particle size distribution, in part due to preventing clump formation and also liberating fine particles from sticking to the grinder. With all other variables kept constant, this shift resulted in at least 15\% higher coffee concentration for espresso extracts prepared from darker roasts. These findings pose financial and sustainability implications, and encourage the widespread implementation of water use to de-electrify coffee during grinding.

\end{abstract}


\begin{keyword}
Coffee \sep Granular electrification \sep Static elimination
\PACS 0000 \sep 1111
\MSC 0000 \sep 1111
\end{keyword}

\end{frontmatter}
\section{Introduction}

Grinding roasted coffee reduces whole beans into flows of highly electrified powders. \cite{lim_triboelectric_2021, mendez2023moisture}  Charged granular materials can lead to electrostatic discharges, \cite{eckhoff2007electrostatic} jamming and sheeting (\textit{i.e.} coating the interior walls of conduits), \cite{eilbeck2000effect, hendrickson2006electrostatics} spontaneous segregation,\cite{mehrotra2007spontaneous} and product non-uniformity.\cite{elajnaf2006electrostatic} Specific to coffee preparation, charging can produce erratic dispersal of grounds, making whole bean grinders somewhat messy. More importantly, however, static charging during grinding results in particle-particle clumping.\cite{lim_triboelectric_2021, mendez2023moisture} These electrostatic agglomerates affect extraction quality when brewing by changing the packing of coffee particles and influencing solid surface area available to percolating water.\cite{cameron_systematically_2020, gagne2020physics} The elimination of these clumps may increase soluble availability by upwards of 15\%, posing substantial financial and sustainability motivations to eliminate their formation. 

\begin{figure*}[!t]
\centering
\makebox[\textwidth][c]{\includegraphics[width=0.9\textwidth]{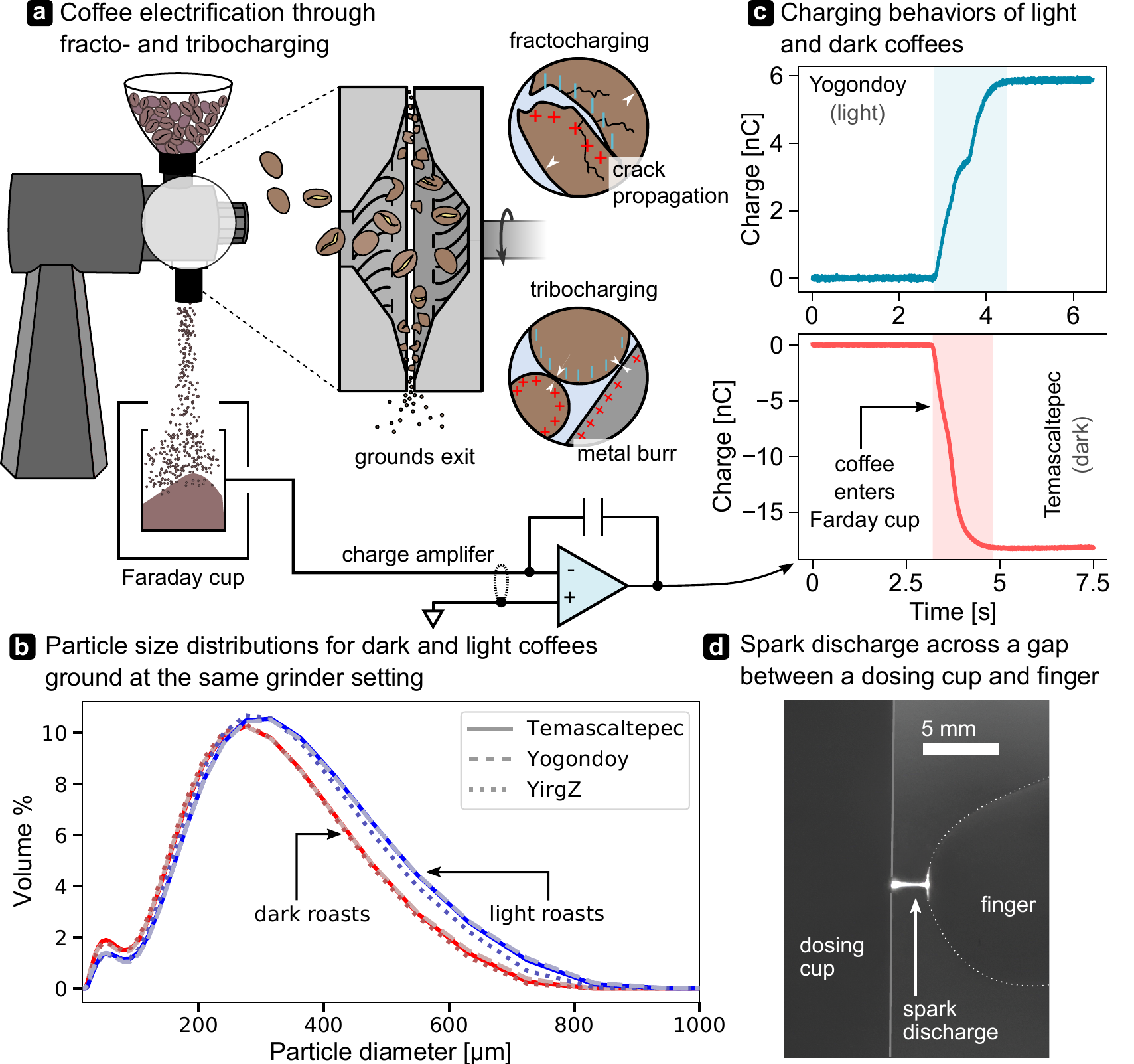}}
\caption{\textbf{Electrification of coffee during grinding. a)} Schematic of the setup used to assess the electrification of coffee during grinding. During fracture, coffee particles accumulate charge
from the burr-coffee and coffee-coffee interfaces (tribocharging), as well as fracture
points (fractocharging). Charge-to-mass ratios can be measured with a Faraday cup and scale. \textbf{b)}Particle size distributions for our in-house roasted coffees ground at setting 2.0 on our Mahlk{\"o}nig EK43. \textbf{c)} Example charging curves (raw data from Faraday cup) for lighter/wetter and darker/dryer coffees. \textbf{d)} Photograph of a spark discharge spanning the gap between a metal cup containing freshly ground coffee and the lead author's finger. Assuming a breakdown field of 3 MV m\textsuperscript{-1} (air at 101 kPa), the potential difference between the two surfaces is $\sim$7.5 kV.}
\label{figure1_schematic}
\end{figure*}

Recently, we demonstrated parameters that control charging by grinding commercially-sourced coffees and measuring the charge-to-mass ratio of the grounds using the process presented in \textbf{Figure \ref{figure1_schematic}a}.\cite{mendez2023moisture} As whole beans are fractured into small grains with broad size distributions (\textbf{Figure \ref{figure1_schematic}b}), \cite{uman2016effect}
particles may acquire charge densities comparable to those of volcanic ash and thundercloud ice, through both fracto- and triboelectrification.\cite{miura2002measurements, mendez2021charge} Generally, the polarity and magnitude of charge loosely depends on the roast level or color of a coffee, and we observed that dark roasts charge largely negative, and lighter roasts charge positively, \textbf{Figure \ref{figure1_schematic}c}. Residual moisture levels, a property typically inversely proportional to color, was found to be the primary determiner of charge polarity, where positive-to-negative charging transition occurs at moisture contents less than $\sim$ 2\%. Some coffees, especially dark coffees, can charge sufficiently to cause gaseous breakdown in the form of millimeter-long spark discharges, \textbf{Figure \ref{figure1_schematic}d}.

The coffee industry has long maintained an intuitive understanding that water can significantly modulate grinding-associated electrostatics. A small amount of water added to whole beans before grinding--the so-called Ross droplet technique (RDT)--is known to prevent static accumulation, and causes the grinder to retain less grounds within its chamber.\cite{homebarista2012static, mendez2023moisture} However, the inclusion of too much water may result in caking or corrosion in the grinder, the limit of which will depend on the coffee and grinder. Concurrently, there is interest in the industry in developing water-free charge mitigation strategies, but their utility and impact on achieving high extraction yields remains unknown. In this paper, we examine the effectiveness of various electrochemical techniques to suppress electrostatic build up -- grinding and waiting, adding external water, and alternate ionization methods. We show that de-electrification techniques that introduce charges after the agglomerates have formed (\textit{i.e.} simply waiting or ionization methods) do not improve the particle availability in espresso brewing, resulting in variable extractions and no improvement in total dissolved solids (\%TDS). The addition of water mitigates charging during the grinding process, and results in extraction increase of beyond 15\%.

\section{Methods}
We electrified coffee by grinding whole beans using a Mahlk\"onig EK43 grinder with stock 98mm burrs. Here we conducted experiments using three coffees (two Mexican coffees and one Ethiopian) roasted in-house on an Ikawa Pro100 roaster following two temperature profiles (see \textbf{Figure S1}), yielding dark and light colors. Salient characteristics of both green and roasted coffee are noted in \textbf{Table \ref{tbl:cafe}}. The coffee color/roast degree and internal water content were measured using The Dipper KN-201 and a RoastRite RM-800, respectively.

We assessed the performance of four techniques to reduce static generated during grinding: time-resolved discharge, the addition of external water, unipolar corona discharge, and balanced corona discharge. With the exception of the grind-and-wait technique, all experiments consisted in applying an electrostatic reduction technique during grinding and measuring the residual charges on particles exiting the grinder using a Faraday cup. Charge-to-mass (Q/m) ratios were then calculated, allowing comparative analysis between the various techniques. The particulars of each method are described in the sections that follow.

Roasted coffee was stored in sealed, evacuated bags and kept at 253 K. Prior to grinding the coffee was allowed to reach equilibrium temperature before unsealing. For all experiments, we used a grind setting of 2.0 (arbitrary units) on the EK43, producing particle size distributions comparable to those shown in \textbf{Figure \ref{figure1_schematic}b}. Particle size measurements were performed on a Malvern Mastersizer 2000 with the solid-particle feed system, Scirocco 2000. Experiments were conducted at 20$\pm3^\circ$C and 35$\pm7$\% RH. Each Faraday cup experiment was conducted a minimum of 3 times. Surface charges were measured using a Keithley 614 electrometer.

\begin{table}[h]
  \caption{\textbf{Characteristics of in-house roasted coffee used in this work.}}
  \label{tbl:cafe}
\centering
\begin{tabular}{ m{1.2cm}m{2cm}m{2cm}m{1.5cm}  }
\hline
& \textbf{Yirgacheffe
Zero Defect} & \textbf{Temascaltepec} & \textbf{Yogondoy} \\
\hline
\rowcolor{LightCyan}
\textbf{Origin} & Ethiopia & M\'exico & M\'exico\\
\textbf{Producer} & Tamrat \newline
Alemayehu & Federico \newline Barrueta & García Luna \\
\rowcolor{LightCyan}
\textbf{Process} & Washed & Washed & Washed \\
\textbf{\%H\textsubscript{2}O}  (initial) & 12.0 & 8.9 & 9.3 \\
\rowcolor{LightCyan}
\textbf{\%H\textsubscript{2}O} (dark) & 1.0 & 1.3 & 1.1 \\
\textbf{\%H\textsubscript{2}O}  (light) & 2.8 & 3.0 & 3.0 \\
 \rowcolor{LightCyan}
\textbf{Agtron} (dark) & 62.1 & 58.4 & 60.2 \\
\textbf{Agtron} (light) & 88.7 & 70.1 & 93.1 \\

\hline
\end{tabular}
\end{table}

\section{Canvasing charge passivation strategies}

\textit{Time-resolved discharge}

\begin{figure*}[!t]
\centering
\makebox[1\textwidth][c]{\includegraphics[width=0.9\textwidth]{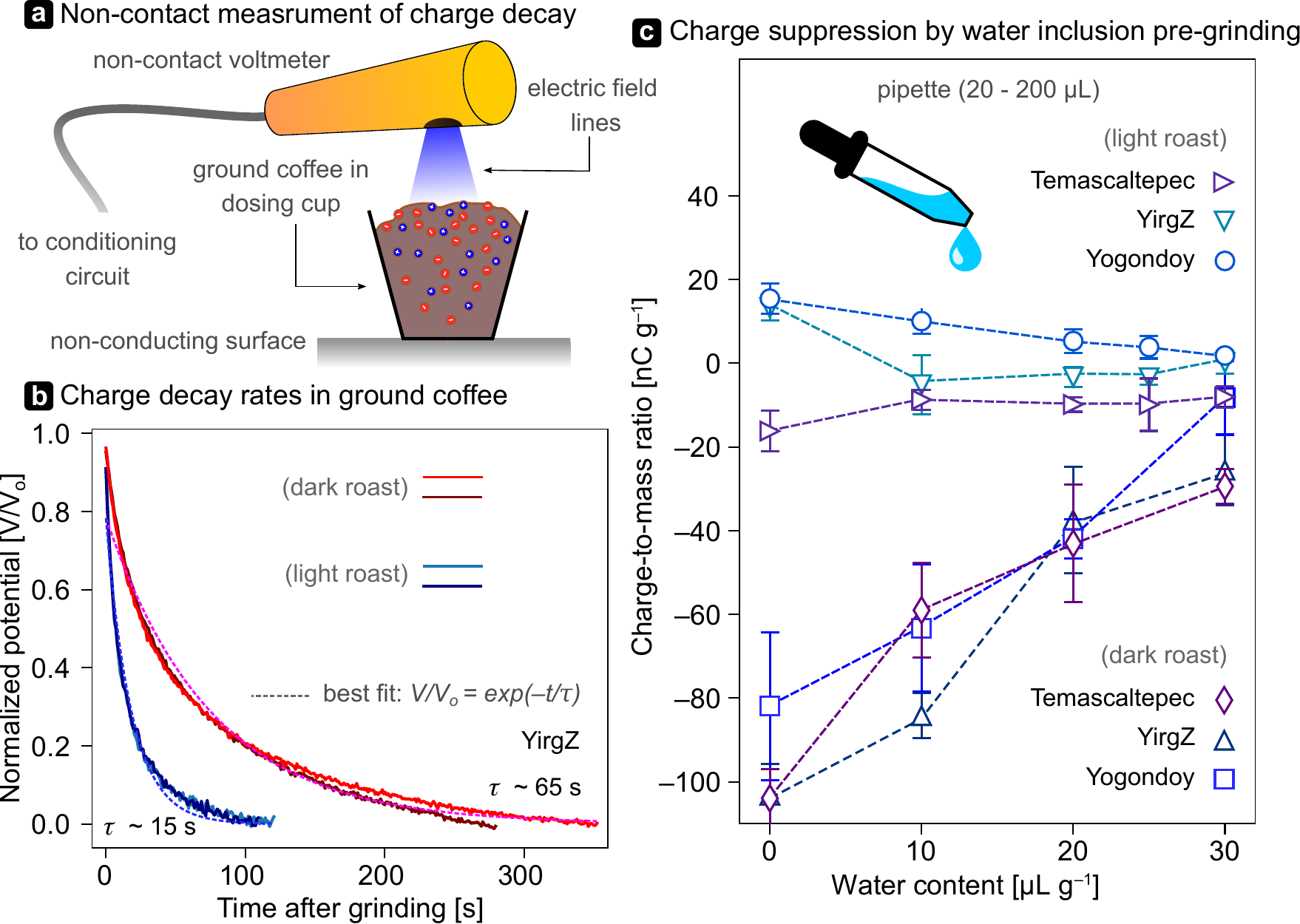}}
\caption{\textbf{Time-resolved and water-mediated charge reduction strategies. a)}  Schematic of setup used to measure the charge decay in freshly ground coffee. A non-contact voltmeter was placed 5 mm above 10 g of ground coffee and its potential was measured every 0.5 sec. \textbf{b)} Charging dissipates with time as charge-carriers recombine with each other through surface and bulk conduction. Some charge may also be lost directly to the atmosphere. Charge in lighter roasts decays faster than in dark coffee. \textbf{c)} During the grinding process, charge accumulation is hindered by the addition of small amounts of water (0 - 30 $\mu$L g\textsuperscript{-1}) to the whole beans prior to grinding.}
\label{figure2_water}
\end{figure*}

Perhaps the simplest discharge method is to let the ground coffee sit for a period of time after grinding. This respite permits discharge through volumetric or surface conduction or, in the case of exceedingly high charges, gaseous breakdown.\cite{paiva2019conduction,  mendez2022lifetime, navarro2023surface} To a zeroth order, Jones and Tang \cite{jones1987charge} have described the relaxation of the volumetric charge density $\rho(t)$ in a powder by; 

\begin{equation}
    \rho(t) = \rho_0 e^{[-t/(\kappa \epsilon_0 \gamma)]}
    \label{relax}
\end{equation}

\noindent 
where  $\rho_0$ is the initial charge density, $t$ is time, $\kappa$ is the dielectric constant of the material, $\epsilon_0$ is the permittivity of free space, and $\gamma$ is the effective resistivity. The denominator in the exponent defines a time constant, $\tau$. This exponential behavior of charge decay can be readily observed using a non-contact electrostatic voltmeter probe (Trek 541A-2) placed 5 mm over 10 g of freshly ground coffee (collected in a metal cup, resting on an insulating surface), \textbf{Figure \ref{figure2_water}a}.  Charge relaxation curves for both light and dark roasts of the same Ethiopian coffee (a washed Yirgacheffe) are rendered in \textbf{Figure \ref{figure2_water}b}. There, the light roast (2.8 \% residual water) loses its charge faster ($\tau \sim$ 15 s) than its darker, drier counterpart (1.0 \% residual water, $\tau \sim$ 65 s). Overall, however, charge appears to dissipate on timescales of minutes, but exceed the average time between preparing shots in a busy cafe. Also, grinding and waiting poses problems for volatile loss and quality degradation, which occurs over similar time frames.\cite{degas}

\subsection{De-electrification through external water inclusion}

While time-resolved de-electrification is both cost-effective and predictable, it does not prevent the formation of aggregates during grinding and also provides the particles with prolonged off-gassing time, resulting in a significant loss of volatiles.\cite{smrke2017time} Thus, a number of active strategies have been devised to address charging during the act of grinding. We recently demonstrated that the addition of extrinsic water mitigated fractocharging in both positive and negative charging coffees.\cite{mendez2023moisture} By incorporating up to 30 $\mu$L g\textsuperscript{-1} of whole coffee beans, our in-house roasts behave similarly to other literature coffee samples, with the charge-to-mass ratio decreasing with increasing water content, \textbf{Figure \ref{figure2_water}c}. Although, in many cases, charging was not completely eradicated, we find that 20-30 $\mu$L g\textsuperscript{-1} reduced the charge by a minimum of 50-60 \%. In practice, this reduction appears to sufficiently mute electrostatic forces, precluding dispersal of grains, clumping, and other effects. A video of the behaviour is presented in the Supplementary Information (\textbf{Video S2}\cite{mendez2023fig}).

Although we have not observed it in our hands, the addition of water could lead to residual water accumulation within the grinder. This could pose problems for bacterial growth within the chamber, corrosion of the burrs, or other effects. By placing a small humidity sensor within the grinding volume of the EK43, we measured the buildup of moisture associated with the RDT. For additive water in the range of 0-50 $\mu$l g\textsuperscript{-1} and a base RH of 40\%, we find that the relative humidity (\%RH) within the grinder may increase up to 75\% for a few seconds (see \textbf{Figure S2}), but returns back to ambient within minutes. The water is presumably consumed in electrochemical reactions, or boiled off. We did not detect condensation. Grinding 10-20 dry beans after grinding the wetted ones returns the grinder to equilibrium \%RH instantaneously.

\subsection{Static reduction using ion beams}

To move away from adding water, charge may also be neutralized by recombination with extrinsically generated negative and positive ions. Such techniques draw from an extensive heritage in other settings,\cite{revel2003generation, pingali2009use,wang2023NASA} and generally employ one of two methods; 1) corona discharge, which uses high-voltages at sharp, conducting tips to accelerate naturally-occurring free ions and also cause collisions with neutrals;\cite{kodama2002static, steinman2006basics, pingali2009use} and 2) ionizing radiation, involving a radioactive or X-ray source to generate similar numbers of positive and negative ions.\cite{czajkowski1999evaluation}. Towards the former, high-voltage ionizers may be unipolar, involving net negative or positive ions, or balanced, where the production of negative and positive charges are equal. Some manufacturers have begun to include these devices in grinders or sell them as accessories to reduce charging.\cite{ionbeam, DF83} Towards the latter, we note that in an earlier version of the manuscript examined de-electrification using a helium nuclei source, Figure S4. However, given the results are largely similar to the effects produced by the balanced ionizer, but balanced ionizer present significantly less risk than using radioactive sources,\cite{CZAJKOWSKI199913} we have elected to not present those data in the manuscript and do not recommend the reader attempt that experiment.

\begin{figure*}[!t]
\centering
{\includegraphics[width=0.9\textwidth]{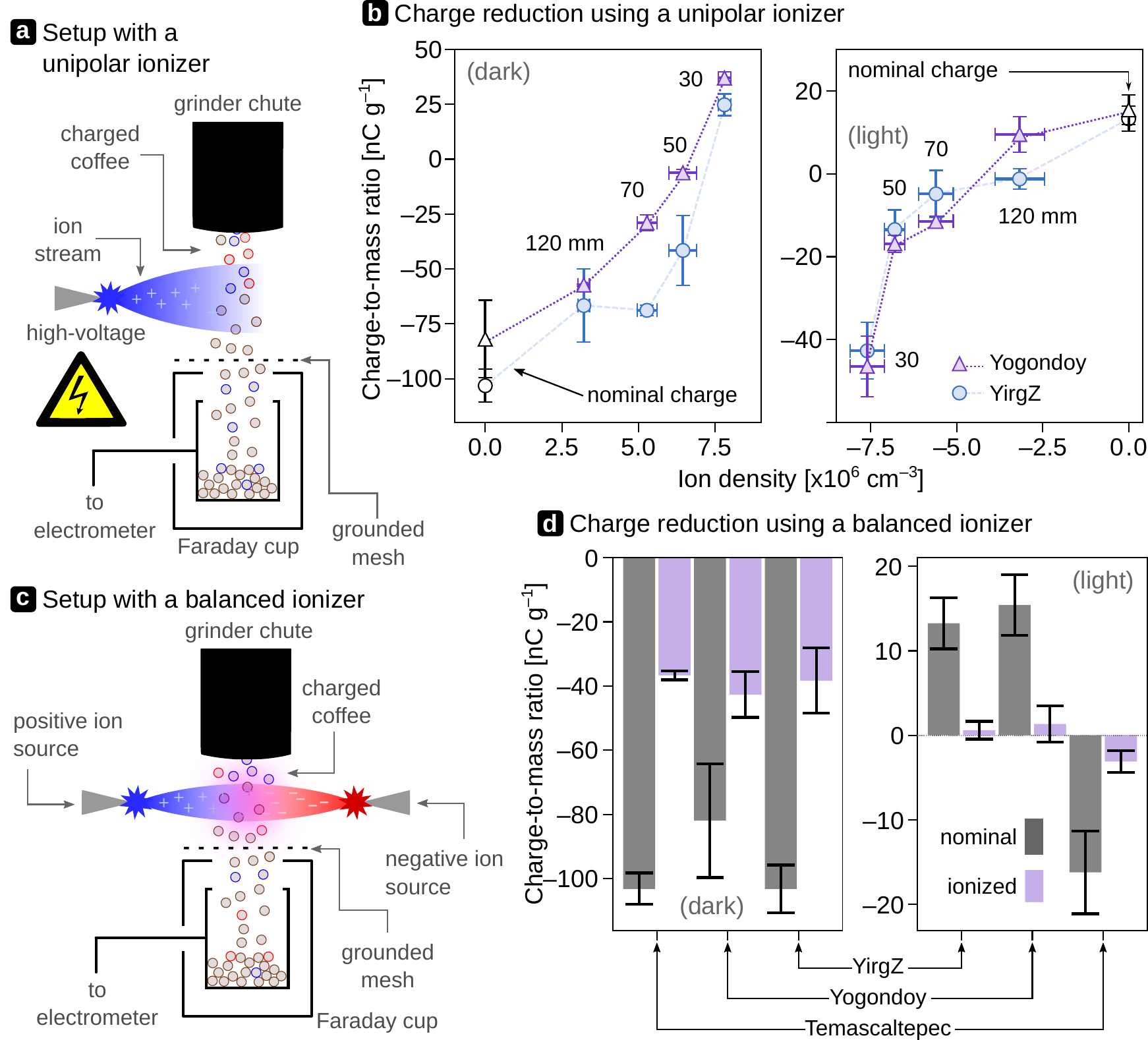}}
\caption{\textbf{Ion beam charge reduction strategies.} Charging may be counteracted by ionizing the air around the coffee grounds as these exit the grinder. Free negative and/or positive ions adhere to solid particle surfaces, tuning their charge. \textbf{a)} Ionization may accomplished via high-voltage gaseous breakdown.  \textbf{b)} While potentially effective, the number of positive and negative ions generated must be adjusted to balance the charging characteristics of a coffee. The nominal charging behavior of coffee with no de-electrification is presented in in white. \textbf{c)} Negative and positive ions may also be generated via a bipolar high-voltage source.  \textbf{d)} Exposing negatively charging coffee to a balanced ionizer reduces its charge by at least 50 \%.}
\label{figure3_ion}
\end{figure*}

\textit{Unipolar ionization} – The efficacy of unipolar ionization on charge reduction can be tested using a high-voltage ionizer. The device consists of a bundle of fine carbon needles fed by a high-voltage source that can generate either 12 or -12 kV. The tip of the ionizer was placed at a controlled distance (30--120 mm) from the coffee grinder chute (see schematic in \textbf{Figure \ref{figure3_ion}b}). Using a Gerdien condenser (AIC, AlphaLabs Inc.), we estimate the negative and positive ion densities to be 1.5 and 1.2 $\times$ 10\textsuperscript{6} cm$^{-3}$ at a distance of 0.3 m, respectively. These densities can be augmented by moving the bundle closer to the chute. To shield the Faraday cup from direct influence of generated ions, we placed a coarse, grounded copper mesh over the cup's aperture. 

The plots in \textbf{Figure \ref{figure3_ion}a} show the Q/m ratios gained by dark and light roasts of the YirgZ and Yogondoy during grinding as a function of ion density, as measured by the Faraday cup. With the ionizer off, the dark roasts nominally charge negatively, whereas the light roasts gain positive charge. Systematically increasing the positive ion density reduces the negative Q/m of the dark roasts toward 0 nC g$^{-1}$ (middle panel of \textbf{Figure \ref{figure3_ion}b}). However, bringing the positive ion source too close to the chute causes a polarity flip and the dark roasts end up depositing positive charge into the Faraday cup. A similar effect is true for light roasts (rightmost panel of \textbf{Figure \ref{figure3_ion}b}); low degrees of negative ionization reduce positive charge, but moderate-to-high ion densities result in negative charging. 

These experiments demonstrate that corona discharge is highly effective at neutralizing charge and minimizing dispersal effects, (\textit{i.e.} mess, see \textbf{Video S3}), but only if the characteristics of the ion source are tuned to the charging behavior of a particular coffee. Despite originating from the same green coffee, the light and dark Yogondoy samples require vastly different ion densities and polarity to achieve a reduction in Q/m ratio comparable to that produced by the water addition technique. The dark roast necessitates ion densities around 6 - 7.5 $\times$ 10\textsuperscript{6} cm$^{-3}$, whereas charge on the light roast is minimized between -5 and -2.5 $\times$ 10\textsuperscript{6} cm$^{-3}$. While these ranges can be achieved by adjusting the distance between coffee and ion source, ion densities outside of these ranges compound existing problems (\textit{increase} charge) and create new problems (such as scattering fine particles via an ionic wind). Because the behavior will depend on environmental variables such as humidity, coffee moisture, roast color, and other parameters, implementing unipolar ionization discharge necessitates trial and error.

\textit{Balanced ionization} – If the charge behavior of a granular material is not known beforehand, implementing electrostatic reduction using ionizing devices that produce an equal number of positive and negative charge carriers may be effective. In general, these bipolar static eliminators concurrently produce positive and negative ions as well as electrons at atmospheric pressure in air. Here, we test a commercially-available bipolar static eliminator (Shopcorp 12M). Like the unipolar setup, the bipolar systems generate ions from bundles of fine carbon needles. The positive and negative outputs were placed facing each other just below the chute of the EK43 grinder with a separation of 10 cm, \textbf{Figure \ref{figure3_ion}c}. In the space between the outputs, we measured positive and negative ion densities in the range of  $\pm$9.8 $\times$ 10\textsuperscript{6} cm$^{-3}$ (for a net density of $\sim$ 0 cm\textsuperscript{-3}). 

During an experiment, ground coffee was allowed to fall through the volume of ionized air generated by the balanced ionizer. The bar graphs in \textbf{Figure \ref{figure3_ion}d} show the Q/m ratios measured on dark and light roasts of all three coffees in the absence (grey bars) and presence (purple bars) of the balanced ionizer. For dark roasts (left panel), we observe a minimum reduction in charge-to-mass ratio of $\sim$ 50\%, comparable to that imparted by the Ross droplet technique (at the highest water contents of 20-30 $\mu$L g\textsuperscript{-1}). For light roasts, we also see up to 90\% reduction in electrification. Interestingly, light-roasted samples that nominally charge positively (YirgZ and Yogondoy) may occasionally acquire small negative charges in the presence of the balanced ionization. We suspect that this polarity flip reflects the higher mobility of electrons relative to that of much more massive positive ions, but it could also depend on the composition of the coffee itself. Note also that the balanced ionizer is overall less efficient than a properly tuned unipolar corona ionizer in its capacity to reduce charge. This observation is aligned with experiments in silos which show that the most effective neutralization of negatively-charged powders occurs not under balanced ionization, but when the positive ion density is higher than the negative one by a factor of 2-3.\cite{zhou2022electrostatic} 

\section{Aggregate formation}
For ionizing strategies, charge mitigation occurs at the chute exit, not in the grinder itself. Conversely, the water addition evidently passivates charging throughout the grinding process. These differences may have important effects on particle-wall adhesion, material loss, and clump formation, especially since previous work has shown that particles adhered to the wall can have Q/m ratios several dozen times larger than those forming the bulk.\cite{zhang2021experimental} 

\begin{figure*}[!t]
\centering
{\includegraphics[width=0.8\textwidth]{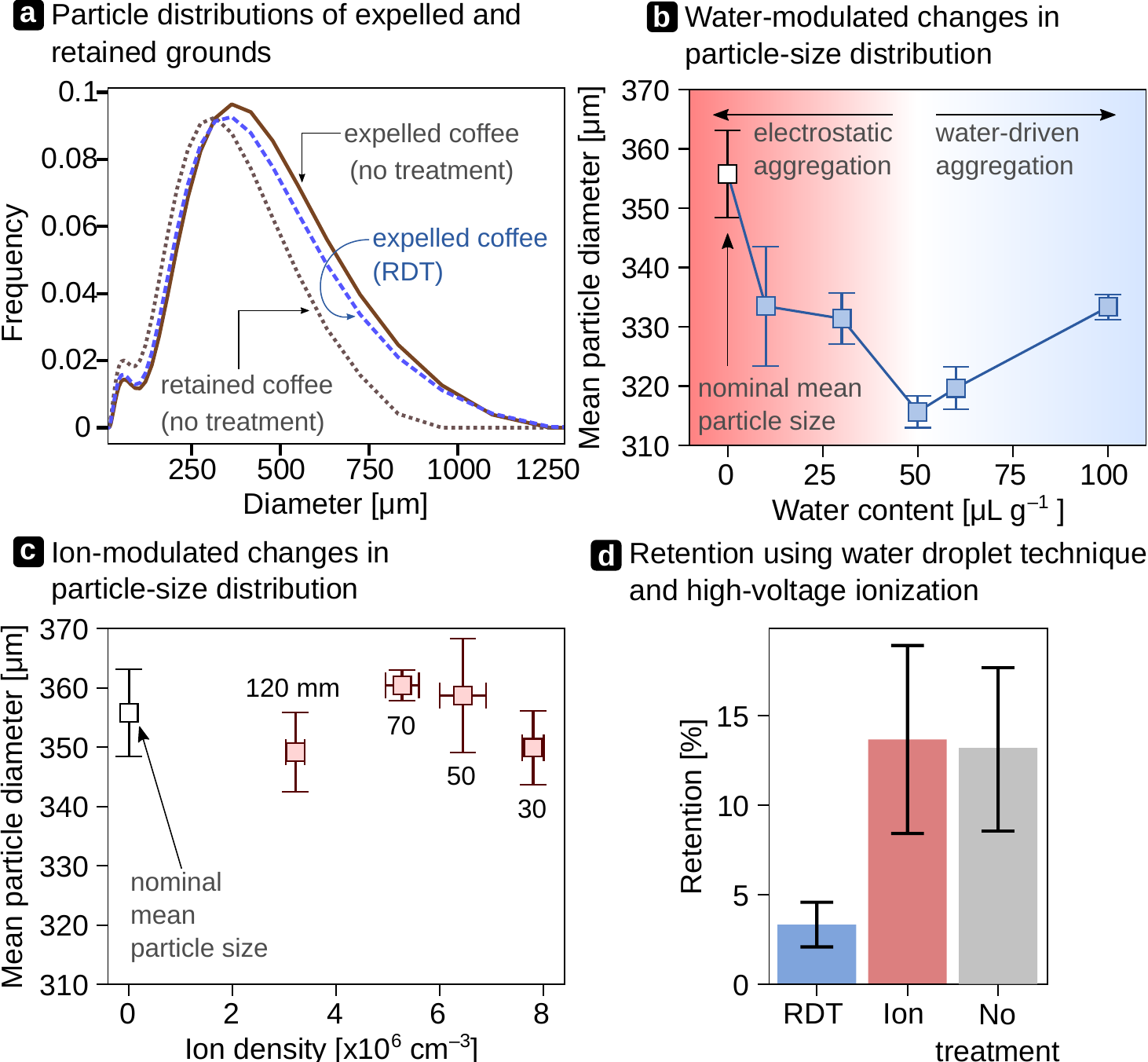}}
\caption{\textbf{Particle aggregation and grinder retention a)} For dry-ground coffee (Yogondoy [dark]), expelled grounds follow the particle size distribution is presented in brown. Grounds retained within the grinding cavity concentrate fines (dotted, brown curve). Fines have higher electrostatic-to-gravitational ratios, meaning they are more likely to adhere to surfaces when charged. Adding even a small amount of water (10 $\mu$l g\textsuperscript{-1}) can significantly reduce electrostatic aggregation, reducing retention and shifting expelled ground particle sizes toward smaller diameters (dashed, blue curve). \textbf{b)} Water contents in the range of 0-50 $\mu$L g\textsuperscript{-1} continue to shift particle sizes toward smaller mean diameters. Water contents above 50 $\mu$L g\textsuperscript{-1} again increases the mean particle size, indicating the activation of wet (capillary) aggregation processes. \textbf{c)} A linear shift in mean particle size and ion density is not observed for coffee treated with a unipolar corona ionizer at different chute-ionizer distances. These data suggest that fine particles within the grinder are not included in the measurement sample (that is, they remain electrostatically adhered to the inner surfaces of the grinder), and the aggregates are formed before deionization, which is to be expected since the corona ionizer is placed after the chute. \textbf{d)} Because the water addition technique (RDT) hinders electrification throughout the grinder, the wet method (using 10 $\mu$L g\textsuperscript{-1}) has the ability to greatly reduce retention. Ionization (7.8 $\times$ 10\textsuperscript{6} cm\textsuperscript{-3}), addressing static only at the grinder chute, involves retention masses similar to those of grinding with no static mitigation treatment.}
\label{figure4_size}
\end{figure*}

Whether a particle of radius $r$ with charge $q$ will electrostatically adhere to a surface depends in part on its electrostatic-to-gravitational force ratio (EGR),

\begin{equation}
EGR = \frac{F_e}{F_g} = \frac{3}{4}\frac{k q^2}{ \pi g \rho_p r^5}.
\end{equation}

\noindent
where $k$ is Coulomb's constant, $\rho_p$ is the particle's density, and $g$ is the acceleration due to gravity. Because $EGR \propto 1/r^5$, smaller particles are much more likely to adhere to surfaces than larger ones (for a given $q$). Thus, dry, charged coffee exiting the grinder tends to be depleted in small particles. Retained fines within the grinder must be knocked out mechanically. This segregation can be readily observed in \textbf{Figure \ref{figure4_size}a}, where we plot the size distribution of grounds directly expelled by the grinder (solid, brown curve) and that of grounds retained within the grinding cavity (dotted, brown curve) when 10 g of whole coffee is ground with no charge mitigation technique.

Our data shows that adding moisture to whole coffee reduces charging by at least 50\%. This reduction would result in a 4-fold decrease in the $EGR$, possibly allowing small particles to overcome electrostatic adhesion and become reincorporated in to the bulk. The effect is reflected in our data where the addition of 100 $\mu$L of water added to 10 g of whole beans produces an appreciable shift in particle size distribution toward smaller diameters, \textbf{Figure \ref{figure4_size}a} (dashed, blue curves). In fact, we observe a decrease in the mean particle size of expelled grounds up to water contents of 50 $\mu$l g\textsuperscript{-1} (see \textbf{Figure \ref{figure4_size}b}). However, at higher extrinsic moisture levels, the mean particle size again moves toward larger diameters due to the formation of moisture-promoted aggregation driven by capillary forces, rather than electrostatic ones.\cite{telling2013ash} 

For the high-voltage ionization system, we do not see an analogous shift toward smaller particles sizes with increasing ion density (see \textbf{Figure \ref{figure4_size}c}). Comparing the masses of coffee retained under high-voltage ionization and the water addition technique, \textbf{Figure \ref{figure4_size}d}, even a modest 10 $\mu$l g\textsuperscript{-1} reduces retention to $\sim$2.5\%, whereas the high-voltage ionization method has retention percentages indistinguishable from those of no-treatment grinding ($\sim$12\%).

\section{The effect of charge mitigation on espresso quality}
In principle, charge mitigation during grinding should provide a better control on the characteristics of the grounds are used during coffee brewing. However, as evident from \textbf{Figure \ref{figure4_size}}, disparate charge reduction methods, while generally effective at reducing a material's Q/m ratio, do not necessarily generate granular materials with equatable properties. How do these differences influence brewing? 

\begin{figure*}[!t]
\centering
{\includegraphics[width=0.8\textwidth]{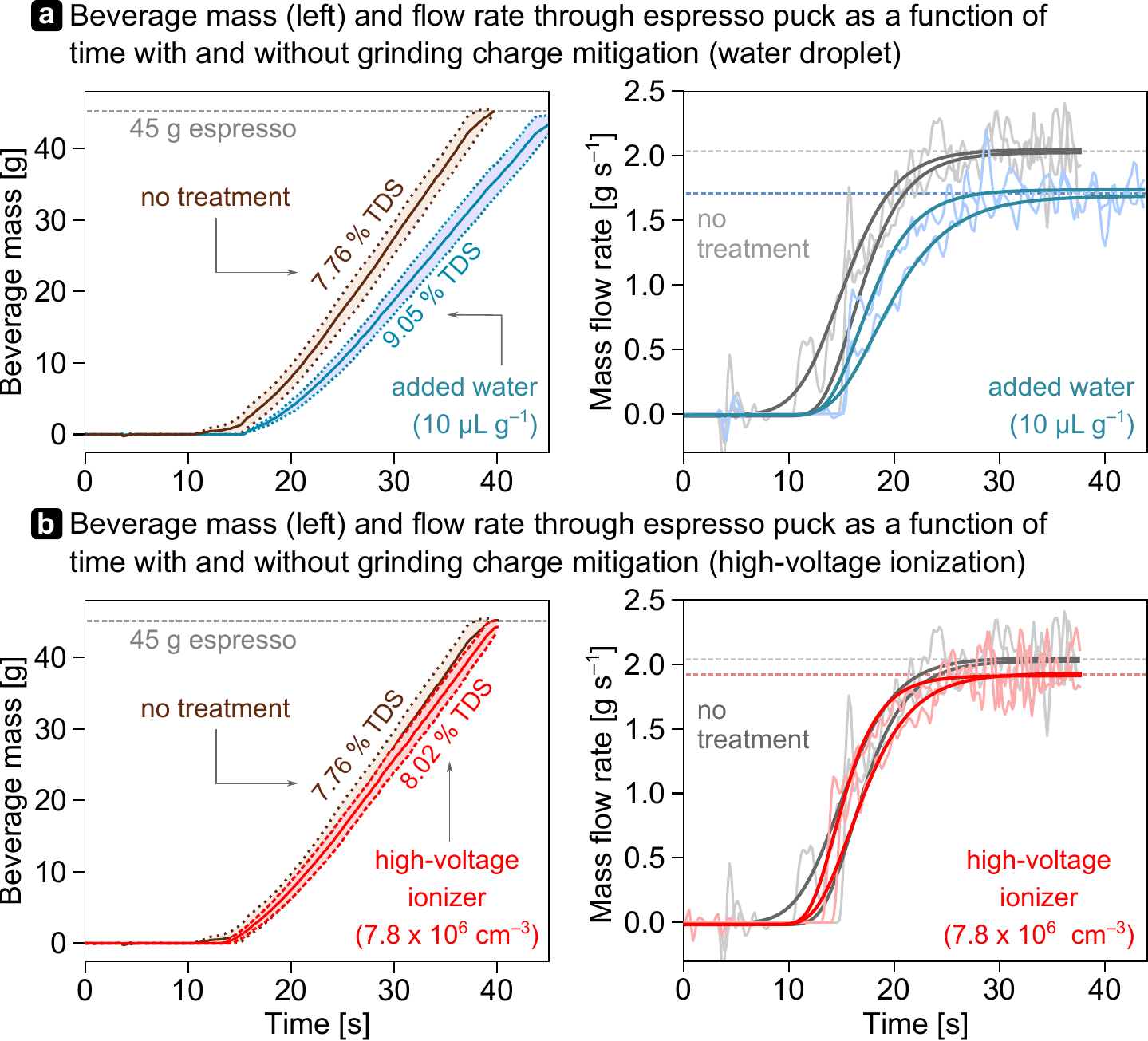}}
\caption{\textbf{Espresso shot time and flow rate dependence with and without charge mitigation for a dark roast (Temascaltepec)} \textbf{a)} Without changing any brewing parameters, coffee prepared using the addition of water to whole beans during grinding produces consistently longer shots (left panel) with reduced flow rates. The shot flow rate can be fit using a generalized logistic function such that the permeability of the bed approaches a constant (right panel). Note that the time it takes an espresso prepared with water to reach this plateau is significantly longer than that of an espresso brewed conventionally. \textbf{b)} Using a positive high-voltage ionizer, we do not observe an appreciable increase in shot time or a reduced flow rate. We do observe a modest increase in \% TDS. The departure from the behavior observed with the Ross droplet technique highlights the fact that ionization methods at the grinder chute do not address electrostatic effects within the grinder cavity and highlights the importance of de-electrification during grinding.}
\label{figure5_ion}
\end{figure*}

In our previous work,\cite{mendez2023moisture} we demonstrated that adding small amounts of water to beans prior to grinding changed the brewing behaviors when preparing espresso. \textbf{Figure \ref{figure5_ion}a} exemplifies such behaviors. There, we plot the shot time (left panel) and flow rate for espresso brewed with and without added water (all other espresso parameters were maintained the same: 18.0 g of dry mass coffee were used to produce 45.0 g of liquid coffee extract, ground at setting 1.0, tamped at 196 N, and brewed using 94 $^o$C water, kept at 7 bar of static water pressure with a 2 second pre-infusion on a Victoria Arduino Black Eagle). For these experiments, we employed the dark roasted Temascaltepec. Mitigating charge with extraneous water results in extended shot times, decreased flow rates, and increased percentages of total dissolved solids (\%TDS) as compared to no charge mitigation. We interpreted these findings to reflect the breaking or rearrangement of electrostatically-bound aggregates, resulting in a particle bed with smaller average grain size and, thus, reduced permeability. The data in \textbf{Figures \ref{figure4_size}a} and \textbf{d} suggest that an additional mechanism may also be operative: added water dislodges a large number of small particles that would otherwise remain trapped within the grinder. 

The impact of shifting the distribution finer is enormous. Keeping all variables the same, a remarkable 16\% increase in coffee concentration is achieved. Such increase in accessible coffee material in comparable brew times, \textbf{Figure \ref{figure5_ion}a}, poses significant financial implications for the coffee industry, allowing for more efficient use of dry mass coffee, at the cost of adding less than 0.5 mL of water to the whole beans during grinding.

Conversely, charge reduction using an ionizing source at the grinder chute does not produce a reduction in the mean particle size, nor does it decrease grinder retention (see \textbf{Figures \ref{figure4_size}c} and \textbf{d}). As evident in \textbf{Figure \ref{figure5_ion}b}, we observe nearly indistinguishable differences in espresso shots prepared with or without the ionizers. The modest increase in \%TDS from 7.76\% to 8.02\% (average over five replicates) is still notable, and can be attributed to some fine particles being liberated from boulders. These data suggests that ionization techniques perhaps do cause some aggregates to break up as these exit the chute, but the effect is limited. However, they do not propitiate the reincorporation of small particles bound electrostatically to the interior surfaces of the grinder and burrs. Thus, unlike the addition of water, discharging coffee at the nozzle produces limited changes to the physical characteristics of the grounds used to brew coffee compared to the untreated samples. 

For dark coffees, added water prior to grinding can generate appreciable differences in espresso brew characteristics, \textbf{Figure \ref{figure5_ion}}. Similar experiments with a lightly roasted sample of the same coffee (Temascaltepec) do not reveal significant changes to espresso brew characteristics, either with extrinsic water or ionization (\textbf{Figure S3}). As noted in our previous work,\cite{mendez2023moisture} darker roasted coffees not only charge negatively, but also acquire the highest absolute charge-to-mass ratios. Lighter roasts charge more ineffectively, with coffees with $\sim$2\% residual water acquiring Q/m ratios near 0 nC g\textsuperscript{-1}. Consequently, electrostatic effects like clumping and sheeting are markedly less present when grinding lighter roasts. That the addition of extrinsic water or ionization do little to modify the properties of the bed is, thus, not surprising. 

\section{Conclusion}
We have assessed the performance of a range of electrostatic reduction techniques in the context of coffee grinding. Added moisture and high-voltage ionization effectively counteract charging generated through fracto- and triboelectric charging. At minimum, both techniques have the potential to decrease the gravimetric charge by at least 50 \%.  Unipolar ionization methods can reduce the Q/m ratios of expelled coffee to near zero if appropriately tuned to a given coffee. Bipolar or balanced ionization may be less effective than unipolar ionization, but can produce charge reductions comparable to those afforded by water addition. However, ionization generally does not address electrification processes and adhesion dynamics within the grinder. As such, ionization methods at the grinder chute do not mitigate material loss (retention) which, for a grinder like the EK43, can be hundreds of milligrams of dry material. Conversely, the charge reduction afforded by water addition (volumes ranging from 0-50 $\mu$L g\textsuperscript{-1} dry mass coffee) has the capacity to resolve aggregation effects across the entire grinding system, even if particles retain some charge. Because particle-wall aggregates comprise smaller grains, we find that reincorporating this material into the bulk significantly changes the espresso brewing behavior (slower shots and smaller flow rates) and resultant increase in coffee solubility. And while our experiments conclusively demonstrate the economic upside of water incorporation during grinding for espresso, we suspect a similar effect will be observed for all percolation brewing techniques, as liberation of fine particles will not only clog void space in the coffee bed, but also fill voids in filter paper in pour-over embodiments.

\section*{Acknowledgements}
JMH acknowledges support from start-up funds at Portland State University. CHH acknowledges the support from the National Science Foundation under Grant No. 2237345 and support from the Camille and Henry Dreyfus Foundation. This work was supported by the Coffee Science Foundation, underwritten by Nuova Simonelli. The work was enabled by donations of green coffee from Farmers Union and Finca La Ilusi\'on. We are grateful to Ikawa for providing the roaster, Pentair for providing the reverse osmosis water filtration system, and Tailored Coffee, Eugene, OR for the use of the EK43.

 \bibliographystyle{elsarticle-num} 
 \bibliography{coffee}







\section*{Supplementary material}

# Supplemental Information: Chemical strategies to mitigate electrostatic charging during coffee grinding

Joshua Méndez Harper*[a], Christopher H. Hendon*[b]

[a]*Electrical and Computer Engineering, Portland State University, 1900 SW 4th Avenue, Portland, 97201, Oregon, US; joshua.mendez@pdx.edu*
[b]*Department of Chemistry, University of Oregon, 1275 University of Oregon, Eugene, 97403, Oregon, US; chendon@uoregon.edu*

## 1. Videos

Three videos are included as part of the supplement. All videos involve ground, dark Temascaltepec coffee, which nominally charges negatively (see **Figure 2**). **Video S1** shows the behavior of ground coffee particles emerging from the EK 43 grinder with no charge mitigation technique. Note the scattering of particles under repulsive (negative) electrostatics forces. **Video S2** shows the behavior using the water droplet technique (at 10 $\mu$l g$^{-1}$). The particle stream becomes much more collimated and dense. **Video S3** shows the behavior using the positive high voltage ionizer (with the tip at a distance of 3 cm from the centerline of the grinder chute; white boom at the right of the frame). While less collimated than the flow with added water, electrostatic forces from charging appear to be greatly reduced by the ionizer. All videos were shot with a Phantom Miro4 high-speed camera and a Canon 100 mm f/2.8 USM lens at 1000 frames per second (25 fps playback).

## 2. Supplementary figures

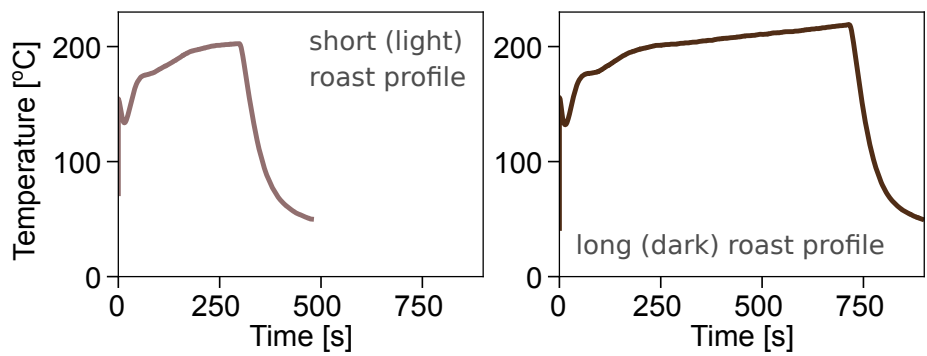

Figure S1: **Temperature profiles used to generate light and dark roasts:** We roasted the coffees listed in **Table 1** using a Ikawa Pro roaster. We roasted coffee in batches of 50 grams. The profiles may be downloaded from 10.6084/m9.figshare.23277320.v1.

2

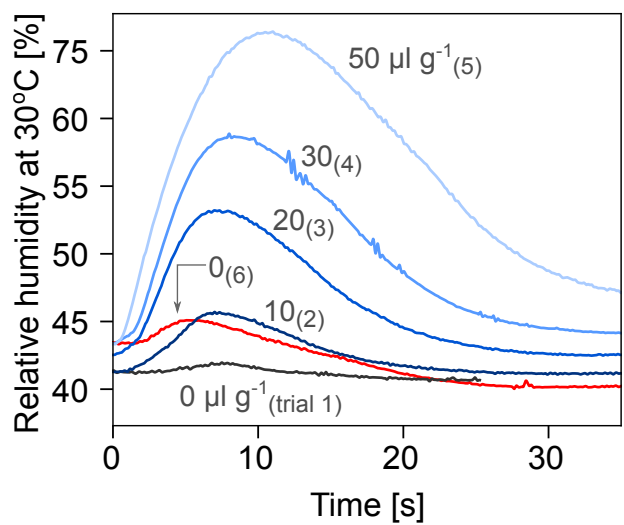

Figure S2: **Changes in humidity within the grinder as function of added water:** A concern associated with the water droplet technique is that moisture could lead to corrosion within the grinder. Thus, we investigated the manner in which small quantities of water changed the environmental conditions inside the grinding cavity. For these experiments, we inserted a small relative humidity (RH) sensor (Honeywell HIH-4030) into the grinder chute and secured it against the stator (non-spinning burr). Relative humidity was then recorded during grinding. Baseline humidity during the experiment was ~40%. The curves rendered above show the variation in RH over time for added water contents in the range of 0-50 $\mu l\ g^{-1}$. Experiments were conducted within minutes of each other in the order indicated by the numbers in parenthesis. For low water contents (~10 $\mu l\ g^{-1}$) we find that humidity in the chamber increases by a few percent and then decays back to background in a few seconds. For higher water contents, RH can increase by up 30%, however we did not register condensation. While the most humid conditions were short-lived, we did find that several minutes were required form the internal RH to return to background for high water contents (>20 $\mu l\ g^{-1}$). This longer-lived excess humidity, however, could be removed effectively by grinding a small amount of dry coffee (red curve, trial 6). Thus, if corrosion is a concern with the water droplet technique, we suggest that a sacrificial, "purge" grind be performed when coffee preparation is concluded.

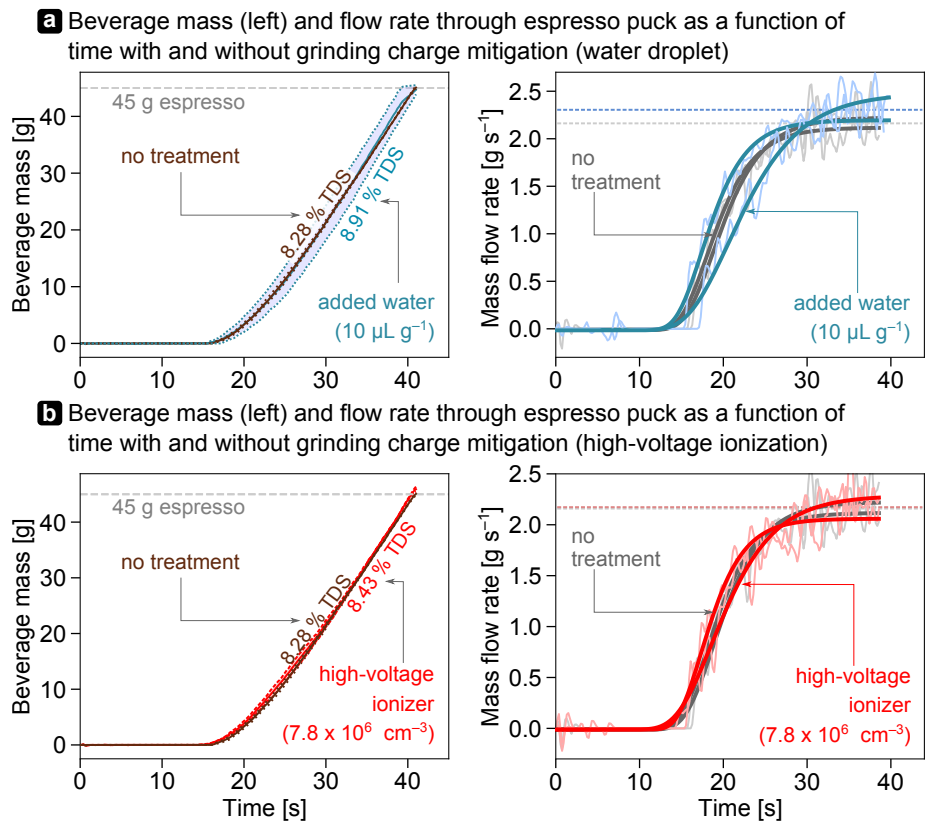

Figure S3: **Espresso shot time and flow rate dependence with and without charge mitigation for a light roast (Temascaltepec) a).** Shot dynamics with (blue curves) and without (brown curves) the addition of extrinsic water. **b)** Shot dynamics with (red curves) and without (brown curves) treatment with high-voltage ionization.

4

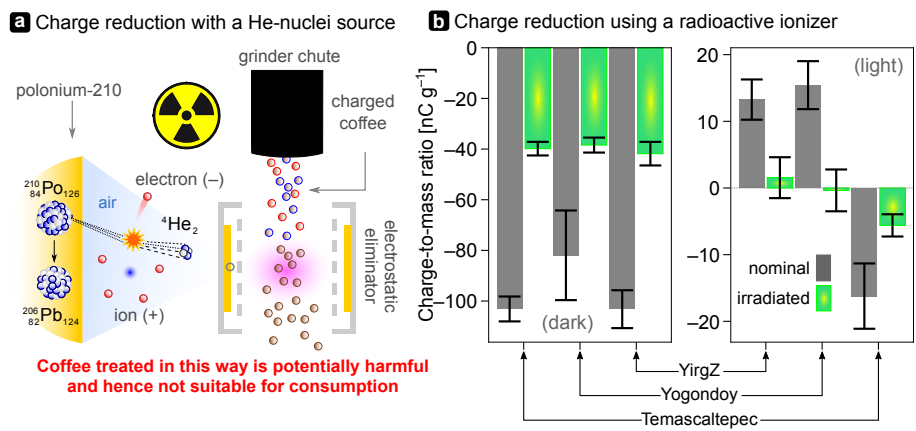

Figure S4: **Although potentially harmful to humans, we also examined de-electrification using helium nuclei. a)** Negative and positive ions may also be generated via alpha decay. **b)** Exposing negatively charging coffee, typically dark roasted, reduces charge comparable to the balanced ionizing method presented in the main text. These experiments were conducted for advancement of basic science and, besides being an inferior de-electrification technique compared to conventional ion beams and/or water addition, the risk associated with the use of radioactive elements makes this approach untenable.



\end{document}